\documentclass[final ,12pt]{clear2024} 

\title[Cautionary Tales on Synthetic Controls in Survival Analyses]{Cautionary Tales on Synthetic Controls in Survival Analyses}
\usepackage{times}
\usepackage{bbm}
\usepackage[capitalise]{cleveref}
\usepackage{wrapfig}
\usepackage[latin1]{inputenc}
\usepackage{algorithm}
\usepackage{algpseudocode}
\newcommand{\squish}[1]{{#1\parfillskip=0pt\par}}
\algblock{Input}{EndInput}
\algnotext{EndInput}
\algblock{Output}{EndOutput}
\algnotext{EndOutput}

\usepackage{graphicx}

\usepackage{hyperref}
\let\svthefootnote\thefootnote
\newcommand\freefootnote[1]{%
  \let\thefootnote\relax%
  \footnotetext{#1}%
  \let\thefootnote\svthefootnote%
}

\newcommand{\indep}{\small {\raisebox{0.05em}{\rotatebox[origin=c]{90}{$\models$}}}}

\clearauthor{%
 \Name{Alicia Curth}\footnote{Work done while an Intern with Microsoft Research.} \Email{amc253@cam.ac.uk}\\
  \addr University of Cambridge
  \AND
  \Name{Hoifung Poon} \Email{hoifung@microsoft.com}\\
  \addr Microsoft Research
  \AND 
  \Name{Aditya Nori} \Email{aditya.nori@microsoft.com}\\
    \addr Microsoft Research
  \AND
\Name{Javier Gonz\'alez} \Email{Gonzalez.Javier@microsoft.com}\\
\addr Microsoft Research
}

\begin{document}

\maketitle

\begin{abstract} Synthetic control (SC) methods have gained rapid popularity in economics recently, where they have been applied in the context of inferring the effects of treatments on standard continuous outcomes assuming linear input-output relations. In medical applications, conversely, \textit{survival} outcomes are often of primary interest, a setup in which both commonly assumed data-generating processes (DGPs) and target parameters are different. In this paper, we therefore investigate whether and when SCs could serve as an alternative to matching methods in survival analyses. We find that, because SCs rely on a linearity assumption, they will generally be biased for the true expected survival time in commonly assumed survival DGPs -- even when taking into account the possibility of linearity on another scale as in accelerated failure time models. Additionally, we find that, because SC units follow distributions with lower variance than real control units, summaries of their distributions, such as survival curves,  will be biased for the parameters of interest in many survival analyses. Nonetheless, we also highlight that using SCs can still improve upon matching whenever the biases described above are outweighed by extrapolation biases exhibited by imperfect matches, and investigate the use of regularization to trade off the shortcomings of both approaches. 
\end{abstract}

\begin{keywords}%
 Synthetic control methods, Matching, Survival analysis, Synthetic control group
\end{keywords}

\section{Introduction}
\squish{The availability of a suitable control group for evaluating the effectiveness of a treatment, policy or other intervention is the backbone of empirical causal inference. Consider the common scenario in which an analyst has access to a small sample of treated instances -- e.g. a single state that experienced a policy shock in economic applications, or a small group that received a novel treatment in an early stage clinical trial -- as well as to a much larger observational sample of potential controls that did not experience the intervention of interest \citep{rosenbaum1985constructing}. A popular approach to use such observational samples to create a control group \textit{tailored to the treated group at hand} has been to rely on \textit{matching} \citep{rubin1973matching, stuart2010matching}: pairing each treated unit with the closest control unit in terms of (a summary of) its observed characteristics. While an intuitively appealing approach, success requires close matches for all units \citep{rosenbaum1985bias}, which becomes challenging  when the dimensionality of characteristics grows only moderately large \citep{stuart2010matching}.}

As an alternative to matching, synthetic control (SC) methods \citep{abadie2003economic, abadie2010synthetic, abadie2021using} have gained rapid popularity in the economic policy evaluation literature recently \citep{athey2017state}. These methods, instead of matching to a single real control unit, construct \textit{synthetic} units that are a weighted average of multiple existing control units, and as such may be able to match the pre-treatment characteristics of the treated units much more closely. Importantly, the literature on these methods has, to the best of our knowledge, focused exclusively on continuous outcomes assumed linear in inputs -- such as GDP per capita \citep{abadie2003economic, pinotti2015economic}, consumption \citep{abadie2010synthetic}, crime rates \citep{donohue2019right, cunningham2018decriminalizing} or stock returns \citep{acemoglu2016value} -- because these are usually of primary interest in economic applications.

The outcome of interest in many medical applications, on the other hand, is a so-called time-to-event or survival outcome \citep{austin2015optimal}, which measures the time elapsed until occurrence of an event (for example, disease progression, an adverse side-effect, or even death). Such survival outcomes are often assumed to follow different data-generating processes (DGPs) than standard continuous outcomes -- instead of linear relationships between characteristics and outcomes directly, relationships are assumed to be more likely on the log-time scale (as in accelerated failure time models \citep{wei1992accelerated}) or log-hazard scale (as in proportional hazards models \citep{cox1972regression}). Further, the target parameters of interest in survival analyses are often also different: instead of only considering differences in expected average survival times due to treatment, other summaries of the survival distribution, relating to the difference of survival curves or ratios of hazards, are of primary interest in many medical studies \citep{clark2003survival}. 

\textbf{Contributions.} In this paper, we  investigate whether and when synthetic controls could serve as an alternative to matching methods in applications where time-to-event outcomes are of interest;  to the best of our knowledge, this is the first work to consider this question. Through our theoretical investigation, we discover multiple obstacles that prohibit the desirable properties that SCs exhibit for standard outcomes to carry over smoothly to the survival context. In particular, we come across two main obstacles: we find that (i) in commonly assumed survival DGPs, SCs will generally be biased for the true expected survival time (even when taking into account the possibility of linearity on another scale in accelerated failure time models) [\Cref{sec:mean-bias}] and (ii) even if they were unbiased for the mean survival time, SCs follow distributions with lower variance than real control units -- which is why other summaries of their distribution, such as survival curves, will usually be biased for the true parameters [\Cref{sec:part2}]. Nonetheless, we also show that using SCs can still improve upon the use of matching whenever the biases incurred above are outweighed by extrapolation biases exhibited by imperfect matches. Finally, we therefore investigate the use of regularization schemes trading off the shortcomings of both approaches [\Cref{sec:reg}] and illustrate how some of the theoretical arguments translate to real data [\Cref{sec:experiments}]. 

{\textit{Remark:} Note that the goal of this paper is not to promote the use of any specific method, but rather to start a discussion on the general applicability of SC-inspired methods in survival analyses due to their importance in medicine. As such, we present first analyses and illustrations of problems arising in this context, with the hope of encouraging future methodological work in this space.}

\section{Background: Synthetic Control in the standard outcome setting}\label{sec:back}
Assume a single individual with pre-treatment characteristics $X_* \in \mathbb{R}^d$ is exposed to a treatment $A_*=1$, and we observe their outcome $Y_*(1) \in \mathbb{R}$; this is the so-called treated unit. We also have access to characteristics $X_{j}$ and outcomes $Y_{j}(0) \in \mathbb{R}$ of $m$ individuals that were not exposed to treatment ($A_j=0$); this is the pool of control units at our disposal. The aim of synthetic control methods is to infer the effect of the treatment by comparing the treated outcome $Y_*(1)$ to that of a \textit{synthetic control unit} $\hat{Y}_*^{SC}$ for the treated unit -- capturing what would have happened to the treated unit had they not been exposed to treatment, i.e. what would have been their \textit{potential} outcome $Y_*(0)$. Such synthetic controls (SCs) are constructed from the outcomes from the control group as a weighted average
\begin{equation}\label{eq:sc}
   \textstyle \hat{Y}^{SC}_* = \sum^m_{j=1} w_j Y_{j}(0)
\end{equation}
where $0\leq w_j \leq 1$ and $\sum^m_{j=1}w_j=1$, with the goal that this synthetic control unit approximates the expected outcome without treatment, $\mu_0(x)=\mathbb{E}[Y(0)|X=x]$, well, i.e. that $\mathbb{E}[\hat{Y}^{SC}_*] \approx \mu_0(X_*)$.

The canonical synthetic control method \citep{abadie2003economic, abadie2010synthetic, abadie2021using} aims to reconstruct the pre-treatment characteristics of the treated units perfectly, i.e. to choose $w=(w_1, \ldots, w_m)$ such that $\sum^m_{j=1}w_jX_{j}=X_*$. As this solution is not always part of the convex hull of observed controls (or even feasible at all), they instead solve
\begin{equation}\label{eq:scobjective}
 \textstyle   \arg \min_{w} || X_* - \sum^m_{j=1}w_jX_{j} ||^2 \textrm{ subject to } 0\leq w_j \leq 1, \sum^m_{j=1}w_j=1
\end{equation}
Throughout, we will refer to synthetic controls that fulfill  $\sum^m_{j=1}w_jX_{j}=X_*$ exactly as \textit{perfect}.

\textbf{Underlying assumptions.} Under a selection on observables assumption ensuring  $Y(0) \indep A | X $ and an overlap assumption ensuring that $X|A=1$ is contained in the support of $X|A=0$, the expected untreated outcome is identified and can be estimated using estimators of the form (\ref{eq:sc}), as discussed in e.g. \cite{kellogg2021combining}\footnote{Recent work \citep{shi2021theory, shi2022assumptions, zeitler2023non} also investigates other identification assumptions for SCs.}. Then, the SC approach of \cref{eq:scobjective} is well-motivated in settings where $\mu_0(x)$ is \textit{linear in} $x$, because then\footnote{\cite{shi2022assumptions} show that linearity assumptions can sometimes be circumvented by instead assuming a more fine-grained model, which, as we discuss in detail in \cref{app:linearity}, is not applicable in the survival setting we study.} we have (for perfect synthetic control units) that
\begin{equation}\label{eq:sc_unb}
 \textstyle   \mathbb{E}[\hat{Y}^{SC}_*] = \sum^m_{j=1} w_j\mathbb{E}[ Y_{j}(0)|X_j]  =  \sum^m_{j=1} w_j \mu_0(X_j) = \mu_0( \sum^m_{j=1} w_j X_j) = \mu_0(X_*)
\end{equation}
Clearly, \cref{eq:sc_unb} holds for linear models $Y_i(0)=X_i\beta + \epsilon$ with mean-zero independent errors $\epsilon$. Further, \cite{abadie2010synthetic} show that when $X$ includes pre-treatment outcomes, \cref{eq:sc_unb} also holds for linear factor models allowing for some shared unobserved factors in the DGP\footnote{The fact that the inclusion of pre-treatment outcomes could correct for some forms of unobserved confounding in this way may be a major contributing factor to the popularity of SC methods in economics.}.

\squish{\textbf{Alternative: Nearest neighbor matching.} A matching estimator can also be written in the form of \cref{eq:sc}, but assigns \textit{all} weight to a single unit $j^*$ satisfying $\arg \min_{j\in [m]}||X_* - X_j||$ -- i.e. the control unit that is $X_*$'s nearest neighbor (NN) in terms of observed characteristics (sometimes also coarsened summaries thereof, e.g. the propensity score \citep{rosenbaum1985constructing}). This outputs    $\textstyle \hat{Y}^{match}_* = Y_{j^*}(0) = \sum^m_{j=1} w_j Y_{j}(0)$ with $w_{j^*}=1$ and $w_{j}=0$ for $j\neq j^*$. Unless there is a perfect match among the control units, we will have that $||X_* - X_{j^*}||>0$ and therefore matching estimators will have to \textit{extrapolate}. In the case of a linear model for a continuous outcome, they would thus incur bias $(X_* - X_{j^*})\beta$ directly proportional to the distance to the match in input space.}

\section{Problem setup: Survival outcomes and analyses}\label{sec:survsetup}
To consider how SC methods could be applied in the survival context in the remainder of the paper, we first introduce the problem setup of interest. Survival analyses are concerned with modelling time-to-event outcomes -- this could be the time $T$ elapsed until death, disease progression or another adverse event occurs.  In addition to estimating standard quantities like the expected potential time $T(A)$ until an event occurs under assignment of different treatments $A$ and given different patient characteristics $X$ (i.e. estimating $\mu_a(x)=\mathbb{E}[T(a)| X=x]$), survival analyses are often interested in modelling quantities capturing more of the \textit{distribution} of survival times (e.g. contrasts of the survival functions $S_a(t;x) = \mathbb{P}(T(a) > t| X=x)$ or hazard functions $h_a(t;x)=\mathbb{P}(T(a) =t|T(a)\geq t, X=x)$) to assess the effects of treatments in this context.

{\textbf{Assumptions.} As in the standard setup discussed above, we assume we have access to a control sample $\{X_{j}, T_{j}(0)\}^m_{j=1}$ (having received treatment $A_j=0$), and rely on the overlap assumption and an unconfoundedness assumption (ensuring here that $T(0) \indep A | X $). Further, we restrict our focus to control time-to-event outcomes generated from commonly considered accelerated failure time (AFT) models which assume event times are generated from models of the form}\begin{equation}\label{eq:aft}
    \log T_i(0) = X_i\beta + \epsilon_i
\end{equation}
{where the $\epsilon_i$ are i.i.d. random variables generated from some common distribution \citep{wei1992accelerated} that then determines the  shape of survival functions. A log-normal AFT has $\epsilon_i\!\sim\!\! \mathcal{N}(0, \sigma^2)$; other popular distributions for $\epsilon_i$ include the logistic distribution and extreme-value distributions, which encompass also the Weibull survival model as a special case.}

\squish{\textit{Remark: Censored outcomes.} In survival analyses, observations are often censored (i.e. lost to follow-up) before any event can be observed. Then, we would only observe $T^{-}\!\!=min(T, C)$ the smaller of event time $T$ and censoring time $C$. For ease of exposition, we disregard censoring in the discussion below. Under the standard, commonly adopted, censoring-at-random (CAR) assumption \citep{laan2003unified}, which requires that censoring and event times are independent when conditioning on observed covariates $X$ and treatment $A$ (or equivalently, that censoring time $C$ is not correlated with the noise term $\epsilon$ in \cref{eq:aft}), the discussion below also directly applies to analyses with censored data, where SC units can then be constructed from uncensored controls only\footnote{To see why this is true, let $\delta=\mathbf{1}\{C < T\}$ be a binary indicator for whether an individual`s survival time was observed; if $\delta=0$ their time was censored.  Then, we can construct synthetic control units from only uncensored control units and write them as $\hat{T}^{SC, \delta=1}_*=\sum^m_{j=1} w^\delta_j T^{-}_j$ where  $w^\delta_j=0$ for all $\delta_j=0$. Note that the CAR assumption implies that $\mathbb{E}[T(0)|X=x, \delta=0]=\mathbb{E}[T(0)|X=x, \delta=1]=\mathbb{E}[T(0)|X=x]=\mu_0(X)$, thus we will have that $E[\hat{T}^{SC, \delta=1}]=\sum_{j: \delta_j = 1} w^\delta_j \mathbb{E}[T(0)|X=X_j, \delta=1] = \sum_{j: \delta_j = 1} w^\delta_j \mu_0(X_j)$. Thus, also in the presence of CAR-censoring, the bias of our synthetic control estimate is determined by the quality of the synthetic match (i.e. the distance of $\sum_{j: \delta_j = 1} w^\delta_j X_j$ from $X_*$) and the linearity of $\mu_0(x)$, as in the absence of censoring.}.}

\textit{Remark: Absence of repeated outcomes.} Unlike economic applications where outcomes are often recorded in repeated panel data structures, outcomes in survival analyses are often terminal (e.g. time to death). Therefore, $X$ would usually not include pre-treatment outcomes in the survival context, and hence SCs cannot account for unobserved factors in the same way they are sometimes assumed to in economic panel models. Note, however, that much of the recent methodological literature on SC methods \citep{abadie2021penalized, kellogg2021combining} considers a general setup identical to ours, differing only in what kind of information could (not) be included in $X$ in spirit.

\section{Obstacles for Synthetic Controls in Survival Analyses (Part 1): Nonlinearity and biases in estimating expected survival time}\label{sec:mean-bias}
In this section, we begin our investigation by considering which biases arise in the canonical synthetic control problem of estimating expected outcomes for a single treated unit $X_*$ as a consequence of switching from simple linear DGPs often assumed for outcomes in economics to time-to-event outcomes generated from AFT models. We then consider issues relating to the distribution of synthetic controls and targeting of other parameters of interest in \Cref{sec:part2}.

\subsection{Interpolation Biases of Synthetic Control in AFT models} 
The AFT assumption implies that $T_i=\exp(X_i\beta+\epsilon)$ -- thus, even for perfect synthetic controls there will likely be substantial bias as $\mu_0(x)$ is not linear in $x$. Why is that? In the context of standard continuous outcomes, \cite{kellogg2021combining} noted that the bias of any estimators of the form \cref{eq:sc} can always be written as
\begin{equation*}
  \textstyle   \mu_0(X_*) -  \sum^m_{j=1} w_j \mu_0(X_j) = \underbrace{[\textstyle\mu_0(X_*) - \mu_0( \sum^m_{j=1} w_j X_j)]}_{\textrm{Extrapolation Bias}}+  \underbrace{[\textstyle \mu_0( \sum^m_{j=1} w_j X_j) -  \sum^m_{j=1} w_j \mu_0(X_j)]}_{\textrm{Interpolation Bias}}
\end{equation*}
Further, \cite{kellogg2021combining} highlight that synthetic control estimators minimize extrapolation biases (as $X_* \approx \sum^m_{j=1} w_j X_j$), but might suffer from interpolation biases if $\mu_0( \sum^m_{j=1} w_j X_j) \neq \sum^m_{j=1} w_j \mu_0(X_j)$ -- i.e.  if $\mu_0(x)$ is nonlinear in $x$. A matching estimator, conversely, which assigns all weight to the single closest control unit $j^*$ and thus uses  $w_{j^*}=1$, will suffer from some extrapolation biases whenever there is no perfect match but has no interpolation bias by construction.

\begin{figure}[t]
\centering
\subfigure[Interpolation bias is high on the nonlinear part of the exponential.]{
    \includegraphics[width=.4\textwidth]{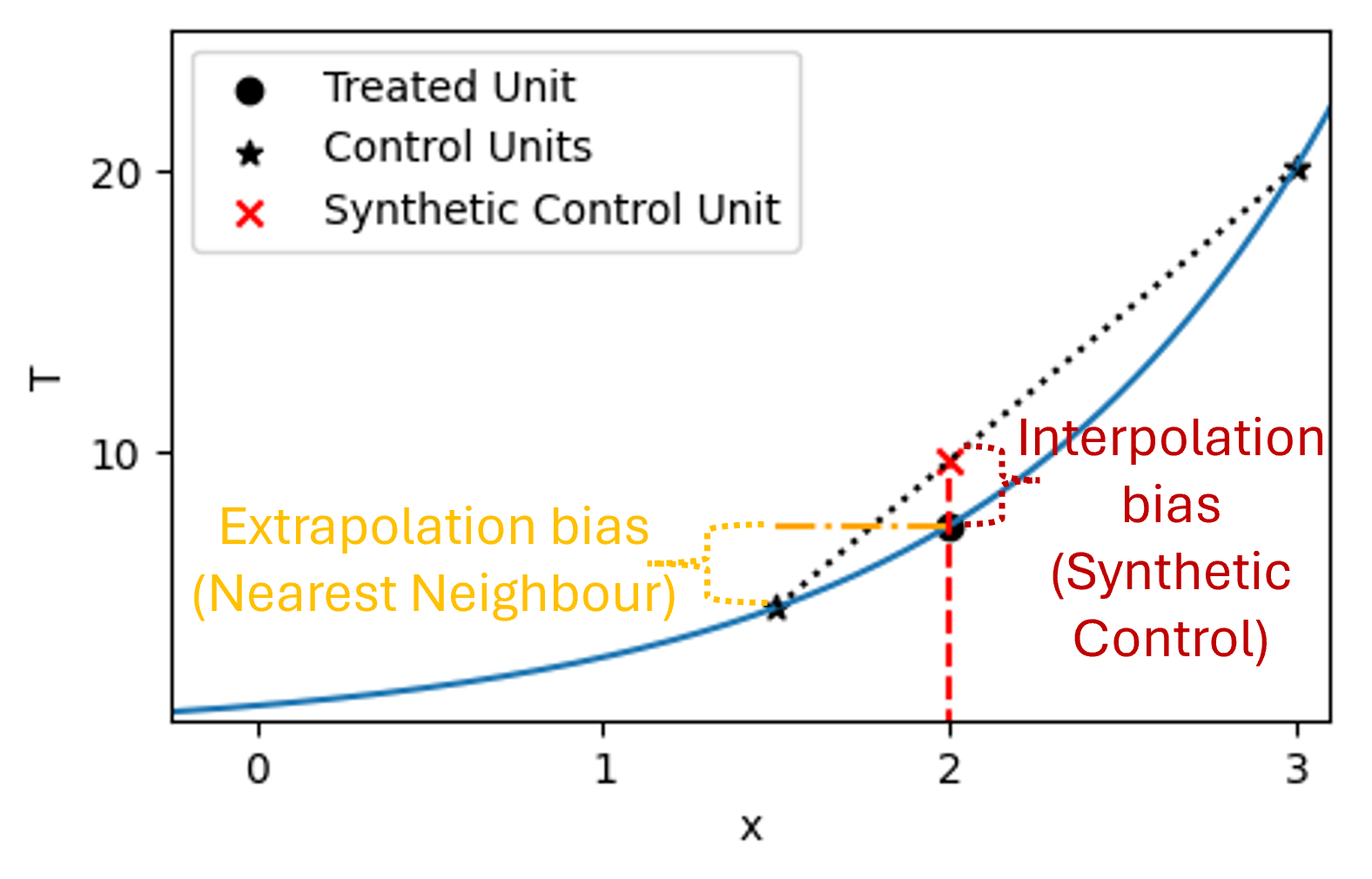}}
    \label{fig:biasbyvar1}
\subfigure[Interpolation bias is negligible on the approximately linear part of the exponential.]{
   \includegraphics[width=.4\textwidth]{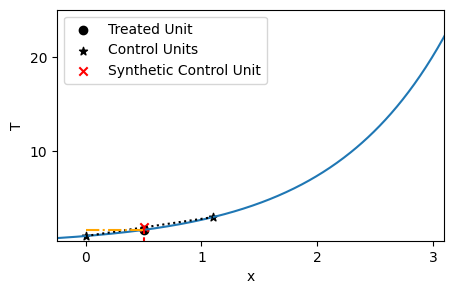}}
    \label{fig:biasbyvar2}       
\caption{\textbf{Stylized illustration of interpolation and extrapolation biases} using a noise-free AFT model ($T=\exp(X)$).}
    \label{fig:interpolation-exp}
\end{figure}

Can we quantify this interpolation bias of perfect synthetic controls in AFT models? Assume the simplest setting where there is no noise, i.e. all $\epsilon_i=0$. Then, $\mathbb{E}[T_*(0)] -\mathbb{E}[\hat{T}^{SC}_*]= \exp(X_*\beta) - \sum^m_{j=1} w_j \exp(X_j \beta) \leq 0$ and synthetic controls will \textit{overestimate} the expected outcome; this bias will be substantial if the $X_j\beta > 1$. This possibility for interpolation bias is also illustrated in \cref{fig:interpolation-exp}, where we see that the relative bias incurred due to interpolation versus extrapolation indeed depends on \textit{where} on the exponential the treated unit and the control units are located. 

\subsection{Does constructing synthetic controls on the  log-time scale overcome this bias?}\label{sec:logsc} It is obvious that the source of interpolation bias here is a result of the AFT assumption: $T$ is \textit{not} linear in $X$ -- but $\log(T)$ \textit{is}! It is thus a natural next step to consider whether we can exploit this assumption, and  apply synthetic control weights to log-time instead. Indeed, for perfect synthetic control weights we have that
\begin{equation}
  \textstyle \widehat{\log(T_*)}^{SC} = \sum^m_{j=1} w_j \log(T_j(0)) =   \sum^m_{j=1} w_j (X_j \beta + \epsilon_j) = X_*  \beta + \sum^m_{j=1} w_j \epsilon_j
\end{equation}
and further, because the $\epsilon_j$ are i.i.d. by assumption and $\sum^m_{j=1} w_j=1$
\begin{equation}\label{eq:aft-logexp}
\begin{split}
  \mathbb{E}[ \widehat{\log(T_*)}^{SC}] = X_*  \beta +     \sum^m_{j=1} w_j   \mathbb{E}[\epsilon_j] =  X_*  \beta + \mathbb{E}[\epsilon_*] \sum^m_{j=1} w_j  =   X_*  \beta + \mathbb{E}[\epsilon_*] = \mathbb{E}[\log(T_*(0)]   
\end{split}
\end{equation}
\squish{Thus, under AFT assumptions, SCs \textit{can} be used to obtain unbiased estimates of \textit{log-survival times}. }

Is this all we needed to make SCs fit for use in survival analyses? Unfortunately, no -- unless log-time is the outcome of interest! If, however, the quantity of interest is the time on the original scale, we might be tempted to simply consider $\hat{T_*}^{log-SC}=\exp( \widehat{\log(T_*)}^{SC} ) = \exp(\sum^m_{j=1}w_j\log(T_j))$. Unfortunately, this will not necessarily lead to unbiased estimates of $T_*(0)$ because for two random variables $A$ and $B$, $\mathbb{E}[A]=\mathbb{E}[B]$ does not generally imply $\mathbb{E}[f(A)] = \mathbb{E}[f(B)]$ for nonlinear functions $f$. In particular, if we were to consider a simple log-normal AFT with $\log(T_i(0)) \sim \mathcal{N}(\mu_i, \sigma^2)$, then $\mathbb{E}[T_i(0)]=\exp(\mu_i+ \frac{\sigma^2}{2})$ -- thus matching the expected value $\mu_i$ of log-time alone clearly is not enough. Instead, more of the distribution -- in this case also the variance -- of true control units would need to be matched. Below, we thus investigate the distribution of synthetic controls further and return to the question of the bias of $\hat{T_*}^{log-SC}$ in \Cref{sec:varbias}.

\section{Obstacles for Synthetic Controls in Survival Analyses (Part 2): Distributional mismatches between synthetic \& real controls and their consequences}\label{sec:part2}
Next, we consider the distribution of synthetic control units beyond just their mean. In \Cref{sec:dist}, we begin by highlighting that synthetic control units will generally have substantially lower variance than real units. In \Cref{sec:varbias}, we show what this means for the bias of synthetic controls constructed on log-scale (as discussed in \Cref{sec:logsc} above). In \Cref{sec:survbias} we then discuss how differences in distribution between synthetic and real units will bias the estimation of survival curves.

\subsection{Aside: On the variance of synthetic controls}\label{sec:dist}
Clearly, \cref{eq:aft-logexp} can be generalized to any model in which an invertible transformation of $f$ of the random variable $Y_i$ follows a linear model $f(Y_i)=X_i\beta+\epsilon_i$ with independent error terms $\epsilon_i$ with equal expected value $\mathbb{E}[\epsilon_i]=\mu_\epsilon$: In any such model, for a perfect synthetic control unit 
\begin{equation}
 \textstyle     \mathbb{E}[\widehat{f(Y_*)}^{SC}]=\mathbb{E}[\sum^m_{j=1}w_j f(Y_j)]=X_*\beta + \mu_\epsilon \sum^m_{j=1}w_j = \mathbb{E}[f(Y_*)]
\end{equation} 

Thus, the expected value can always be matched on the $f$-scale. What about the variance of such perfect synthetic controls? As synthetic controls are simply weighted averages, it is easy to see that, for independently generated homoskedastic error-terms with variance $\sigma_\epsilon^2$,
\begin{equation}\label{eq:SCVAR}
  \textstyle    \text{Var}(\widehat{f(Y_*)}^{SC}) = \text{Var}(\sum^m_{j=1}w_j \epsilon_j) = \sigma_\epsilon^2 \sum^m_{j=1}w_j^2
\end{equation}
Further, because synthetic control weights are restricted to the convex hull of the data (thus all $0\leq w_j \leq 1$ and $\sum^m_{j=1}w_j=1$), we have that $\min(\sum^m_{j=1}w_j^2)=\frac{1}{m}$ and $\max(\sum^m_{j=1}w_j^2)=1$, which are attained at $w_j=\frac{1}{m}$ (the sample average) and $w_j=1\{j=k\}$ (for $k$ the index of a single control unit), respectively. Thus, $    \text{Var}(\widehat{f(Y_*)}^{SC}) \leq \text{Var}(f(Y_*))$, and the variance of synthetic control units $\widehat{f(Y_*)}^{SC}$ is strictly smaller than that of real outcomes $f(Y_*)$ whenever there is more than one contributor (which would generally be the case unless there is a perfect match for $X_*$ in the control sample). Note that his observation holds generally for synthetic controls,  including the standard synthetic control setting where $f$ is the identity function.

\subsection{Bias of log-time synthetic controls for the expected survival time in log-normal AFT models (ct'd from \cref{sec:logsc})}\label{sec:varbias}
For the special case of log-normal AFT models, we can use \cref{eq:SCVAR} to deduce the bias of perfect synthetic controls $\hat{T_*}^{log-SC}=\exp( \widehat{\log(T_*)}^{SC})$ constructed on log-scale. In particular, due to the properties of the normal distribution, we know that $ \widehat{\log(T_*)}^{SC}\sim \mathcal{N}(\mu_*,\sigma_\epsilon^2 \sum^m_{j=1}w_j^2)$, where $\mu_*\!\!=\!\!\mathbb{E}[\log(T_*)]\!\!=\!\!X_*\beta + \mathbb{E}[\epsilon]$. Therefore, using properties of log-normals,  $\mathbb{E}[\hat{T_*}^{log-SC}]\!\!=\!\!\exp(\mu_* + \frac{\sigma_\epsilon^2}{2} \sum^m_{j=1}w_j^2)$, which is a biased estimator for $T_*(0)$ whenever there is more than one contributor because then $\sum^m_{j=1}w_j^2<1$ while the true expected value of $T_*(0)$ is $\mathbb{E}[T_*(0)]=\exp(\mu_* + \frac{\sigma_\epsilon^2}{2})$.\footnote{Note that, if time is assumed log-normal and $\sigma_\epsilon$ can be estimated or is known, $\hat{T}^{log-SC}_*$ could therefore in principle be debiased by multiplying it by a factor $\frac{\exp(\sigma_\epsilon^2/2)}{\exp(\sigma_\epsilon^2/2 \sum^m_{j=1} w_j)}$. We do not investigate this avenue further because usability of this specific debiased estimator in practice requires both i) availability of a perfect synthetic control unit (as otherwise the mean $\mu_*$ is also not matched) and ii) log-\textit{normality} of the true time (as different AFT models would require different correction factors depending on the underlying relationship between the distributions of log-time and time), which is a very strong assumption and unlikely to hold in many applications.}

\begin{figure}[t]
\subfigure[$\sigma=0$]{
    \includegraphics[width=.3\textwidth]{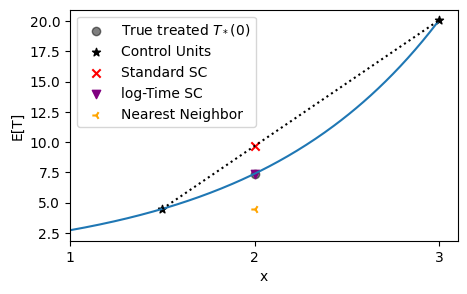}}
    \label{fig:sig0}
\subfigure[$\sigma=1$]{
   \includegraphics[width=.3\textwidth]{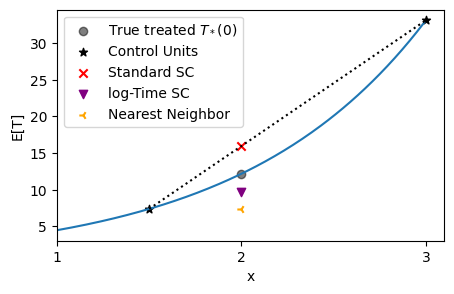}}
    \label{fig:sig1}   
    \subfigure[$\sigma=2.5$]{
   \includegraphics[width=.3\textwidth]{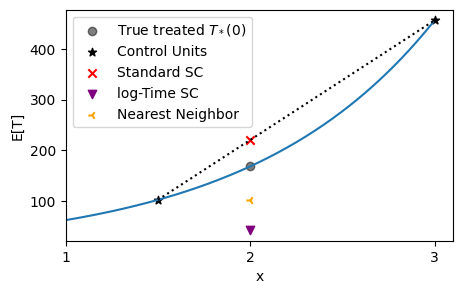}}
\caption{\textbf{Illustration of the effect of error variance on bias of different estimators} for a log-normal AFT with $\log(T)=X + \epsilon$ and $\epsilon\sim \mathcal{N}(0, \sigma)$.}
    \label{fig:varbias}\vspace{-1cm}
\end{figure}

The estimator $\hat{T_*}^{log-SC}=\exp( \widehat{\log(T_*)}^{SC})$ will thus generally \textit{underestimate} the expected (mean) survival time, and this bias will be more severe a) the larger the error variance is  relative to $\mu_*$ (i.e. the lower the signal-to-noise ratio) and b) the more contributors there are (this is because $ \sum^m_{j=1}w_j^2$ becomes smaller as the $w_j$ become more uniform and \textit{less sparse}). We illustrate the bias induced by variance-mismatches for a simple log-normal AFT in \cref{fig:varbias}.

In  \cref{fig:varbias}, we also observe that the relative performance of a nearest neighbor match and the standard synthetic control is unaffected by the error variance, while the relative performance of synthetic control constructed at log-time scale greatly deteriorates as the variance increases. Further insight on this special case can therefore be gained by considering explicit expressions for the bias:

\begin{align}
  \textstyle        \mathbb{E}[T_*(0)] -\mathbb{E}[\exp(\widehat{\log(T_*)}^{SC})] = \underbrace{ \textstyle \left(\exp(\frac{\sigma^2}{2}) - \exp(\frac{\sigma^2\sum^m_{j=1}w_j^2}{2})\right)}_{\text{always }\geq 0}\exp(X_*\beta) \\
    \mathbb{E}[T_*(0)] -\mathbb{E}[\hat{T}^{SC}_*] =  \exp(\frac{\sigma^2}{2})\underbrace{\left(\exp(X_*\beta) -  \textstyle \sum^m_{j=1}w_j\exp(X_j\beta)\right)}_{\text{always } \leq 0 \text{ as } \exp(\cdot) \text{ is convex }} \\
            \mathbb{E}[T_*(0)] -\mathbb{E}[\hat{T}^{match}_*] = \exp(\frac{\sigma^2}{2})\underbrace{\left(\exp(X_*\beta) - \exp(X_{match}\beta)\right)}_{\leq \geq 0 \text{ depending on exponents}}
\end{align}

Thus, for log-normal AFTs, a standard synthetic control estimator will \textit{overestimate} the expected time while a synthetic control constructed at log-time scale will \textit{underestimate} the expected time. Further, the relative performance of standard synthetic control and matching is determined by the distance of the individual selected control units from the treated unit $X_*$, while the bias of a (perfect) synthetic control constructed at log-time scale is entirely driven by differences in variance due to multiple contributors. More generally, regardless of whether $T(0)$ actually follows any AFT model, we always have $\hat{T_*}^{log-SC}\!\! \!\!= \!\!\exp( \sum^m_{j=1} w_j \log(T_j(0))) \leq \exp( \log \sum^m_{j=1} w_j (T_j(0))) = \hat{T}^{SC}_* $ because of concavity of the $\log$. That is, a log-time synthetic control will always output an estimate of time that is lower than the standard synthetic control estimate.

\textit{Remark: Over- vs underestimation.} Comparing standard and log-time synthetic controls, a natural question that arises is thus whether one would prefer an estimator that overestimates or one that underestimates expected survival time. While neither is ideal, the answer may be application-dependent: if the observational bias one is trying to correct for is that the control sample overall has lower outcomes already (e.g. because less healthy individuals are less likely to receive a promising yet invasive treatment), using an estimator that may systematically underestimate time will likely exacerbate this bias and may therefore be less desirable (and vice versa).

\subsection{Bias in survival curve estimation}\label{sec:survbias} 
Finally, we now consider biases in estimating other summaries of the survival distribution for a control sample -- e.g. marginal survival curves $S_0(t) =\mathbb{P}(T_*(0)>t)$ (marginalised over the distribution of covariates $X_{*}$ among the treated). To do so, we move beyond finding matches for single units and consider creation of an entire synthetic control \textit{group} by creating individual synthetic control units for the outcomes $T_{k*}(0)$ of an entire cohort of $n$ treated units for which we have access to $\{X_{k*}, T_{k*}(1)\}^n_{k=1}$. This setup is motivated by, for example, the problem of constructing a virtual (external) control arm for a one-armed clinical trial \citep{thorlund2020synthetic, mishra2022external}, which is often tackled either by matching or by weighting observed control units in downstream analyses. Here, we are therefore interested in understanding whether one could create synthetic units to be used in such downstream analyses instead.

\squish{\paragraph{Exaggerated concentration around the mean survival time.} To illustrate challenges with this approach even in the simplest of settings, we now assume that time indeed follows a simple linear model $T_i(0)=X_i\beta + \epsilon_i$ so that even standard synthetic control is unbiased for the expected (mean) survival time. However, when the goal is to estimate entire survival functions from the constructed cohorts, the full implied distribution of survival times -- and not just their means -- matters. In particular, because -- as we have shown in \cref{sec:dist} --  the variance of synthetic control units is smaller than that of real units, the synthetic survival times will be \textit{too concentrated around their mean}, which means that survival curves constructed from such synthetic cohorts will likely both i) underestimate the probability of early events (i.e. underestimate $P(T_*(0)<t_1)$ for $t_1$  small) and ii) underestimate the probability of surviving for long times (i.e. underestimate  $P(T_*(0)>t_2)$ for $t_2$  large). }

\begin{wrapfigure}{r}{0.35\textwidth}
\vskip -0.1in
    \centering
    \includegraphics[width=0.35\textwidth]{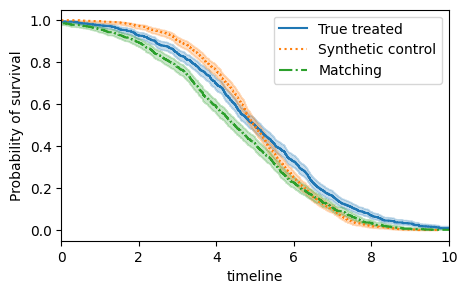}\vspace{-1cm}
    \caption{\textbf{Stylized illustration of biases in survival curve estimation.}} 
 \label{fig:survillustration}\vspace{-.5cm}
\end{wrapfigure}We illustrate this for a very simple example in \cref{fig:survillustration}. We revisit the stylized setup from the previous figures with $x_*=2$ and two control units $x_1=1.5$ and $x_2=3$ but let survival time be linear $T(0)= 3 + x + \epsilon, \epsilon\sim \mathcal{N}(0, 2^2)$ to illustrate that issues arise even in this setting. We generate 1000 examples at each datapoint, and can thus construct the survival curves from 1000 mutually independent nearest neighbor matches and synthetic control units, respectively. \cref{fig:survillustration} clearly illustrates the anticipated issues with the survival curve of the synthetic control cohort ({orange}, dotted): while the mean of the true survival times ({blue}, solid) is matched well, the estimated survival curve does not perform well for low and high survival times. The survival curve created through matching ({green}, dashed), on the other hand, is shifted to the left due to imperfect matching but captures the shape of the true survival curve well. 

\paragraph{Dependence between synthetic control units.} A final issue that arises once we move from individual synthetic control units to constructing a full control group to be used in downstream analyses is that -- if some synthetic units share donors -- they are no longer independent. In particular, for any two synthetic control units we have
\begin{equation}\label{eq:cov}
  \textstyle    \text{Cov}(\hat{Y}^{SC}_{k*}, \hat{Y}^{SC}_{l*}) = \text{Cov}(\sum^m_{j=1} w_{jk} Y_{j}(0), \sum^m_{j=1} w_{jl} Y_{j}(0)) = \sigma^2  \sum^m_{j=1}  w_{jk} w_{jl}
\end{equation}
Thus, the larger $\sum^m_{j=1}  w_{jk} w_{jl}$, the higher the correlation between synthetic control units, which will also lead to \textit{underestimation} of the spread of the survival distribution (as survival times will be artificially close across units). Note that even matching estimators -- which have only a single contributor $j$ that gets weight $w_{jk}=1$ (and hence do not underestimate the individual variance) -- can encounter this issue if matches are chosen with replacement \citep{stuart2010matching}. 

\section{Towards Overcoming Obstacles: Penalizing deviations from real distributions}\label{sec:reg}
In the preceding sections, we highlighted that the use of synthetic controls for survival analyses can lead to bias because their distribution will likely differ from real control units -- depending on the behaviour of the weight vectors $\mathbf{w}_{k}=(w_{1k}, \ldots, w_{mk})$ for each synthetic unit $k$. Motivated by the discussions above, it is therefore a natural next step to consider whether one can trade off the quality of the synthetic match with implied deviations from the true distribution by \textit{penalizing} the $\mathbf{w}_{k}$. 

Here, we consider doing so by trading off the fit in input space (i.e. extrapolation bias) with bias incurred due to \textit{artificially low variance of synthetic control units} by \textit{maximising $||\mathbf{w}_*||^2$} (or equivalently, minimizing $-||\mathbf{w}_*||^2$) which maximises the variance of the synthetic control outcome. For a single synthetic unit, this gives rise to the following penalized objective:
\begin{equation}
  \textstyle    \mathbf{w}_*=  \arg\min_{w: 0 \leq w_i \leq 1, \sum_i w_i =1} ||X_* - \sum^m_{j=1} w_j X_j||^2 - \lambda_{var} ||\mathbf{w}||^2
\end{equation}
Note that -- because $||\mathbf{w}_*||^2$ is maximised when all weight is given to \textit{a single donor unit} -- this objective essentially interpolates between a perfect synthetic control (when $\lambda_{var}=0$) and a nearest neighbor matching estimator (when $\lambda_{var}\rightarrow \infty$) at its extremes\footnote{This behaviour is similar to the penalized objectives of \cite{kellogg2021combining} and \cite{abadie2021penalized}, which are derived with the explicit goal of interpolating between nearest neighbor and synthetic control estimator and therefore use a penalty term $+\lambda \sum^m_{j=1}w_j||X_* - X_j||^2$ instead of our term $- \lambda_{var} ||\mathbf{w}||^2$ that is motivated from concerns about variance.}. Because $||\mathbf{w}_*||^2$ grows as weight is assigned less uniformly, this objective may encourage \textit{more sparsity} in the weights at intermediate values -- which, as a side effect, thus also leads to \textit{less interpolation bias}. While we do not investigate this further here, we note that if the goal was to also correct for (i.e. minimize) the covariance between units, \cref{eq:cov} suggests that this can be achieved by minimizing $\mathbf{w}_k ^T \mathbf{w}_l$ for $l\neq k$, i.e. adding a term $+ \lambda_{cov} \sum^n_{k=1}\sum_{l>k} \mathbf{w}_k ^T \mathbf{w}_l$, which will reduce overlap between donor units. 

\textit{Remark: Initialisation.} Whenever the size of the donor pool is large relative to the number of features, the minimizer of the synthetic control objective will not necessarily be unique. In our experiments (see \cref{fig:initbias}), we found that initialising $\mathbf{w}$ at the nearest neighbor solution leads to much better performance than initialising it with random donors; this initialisation scheme also naturally encourages solutions with large $||\mathbf{w}_*||^2$.

{\textit{Remark: Hyperparameter choice.} Hyperparameters in synthetic control methods are usually chosen by cross-validation on i) pre-intervention outcomes of the treated or ii) post-intervention outcomes of the untreated \citep{abadie2021penalized}. In our context, where no pre-treatment outcomes are available, only option ii) is feasible. We would thus suggest choosing hyperparameters through performing cross-validation \textit{among the controls only} by constructing synthetic control outcomes for held-out out controls which can then be compared to their real outcomes. The used validation metric should depend on the goal of the downstream  analysis: this could either be prediction metrics (e.g. the mean absolute error as in our experiments) if getting the individual survival times right is of greater importance, or metrics capturing the realism of the distribution of survival times (e.g. a Kolmogorov-Smirnov statistic as in our experiments) if that aligns better with the final goal of the analysis.}

\section{Empirical illustrations}\label{sec:experiments}
\begin{wrapfigure}{r}{0.33\textwidth}
\vskip -0.6in
    \centering
\includegraphics[width=0.33\textwidth]{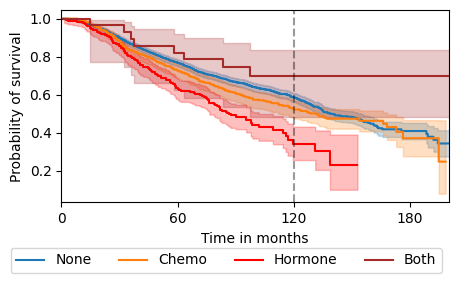}\vspace{-.2cm}
    \caption{\textbf{Kaplan-Meier Survival Curves by Treatment.}} \vspace{-5mm}
 \label{fig:underlyingsurv}
\end{wrapfigure}Finally, we illustrate some of the theoretical discussion presented above on real data. We use the Rotterdam breast cancer dataset \citep{foekens2000urokinase, royston2013external}\footnote{Retrieved from the \texttt{R} package \texttt{survival} \citep{survival-package}.}, which contains 2982 female breast cancer patients of the Rotterdam tumour bank. It provides information on patient characteristics (age, menopausal status, tumor size, tumor differentiation grades, number of positive lymph nodes, progesterone receptors and estrogen receptors), information on treatment (whether hormone therapy and/or chemotherapy was received) and information on a survival outcome (the time from primary surgery to death or censoring).  We plot the marginal survival curves by treatment in \cref{fig:underlyingsurv}. Throughout, we focus on survival within the first 10 years to ensure sufficient samples with follow-up.

\subsection{Experiment 1: Evaluating relative performance through biased resampling of real data}\label{sec:exp1}
{In this section, we use the data of the $n=2091$ patients that received neither treatment to evaluate the performance of the different approaches. Inspired by negative control outcome methods \citep{lipsitch2010negative, shi2020selective}, we create biased subsets (``treatment groups'') of this data and evaluate methods based on their ability to reduce bias that we induced ourselves in  a setting where we know there is \textit{no effect at all} (all patients actually received the same treatment!). In particular, as we detail further in \cref{app:exp}, we sample a ``target group'' of size $\approx .05n$ with \textit{higher} expected survival time (estimated using a Cox model) than the average patient. While unadjusted comparisons will thus find better survival in this target group compared to the remaining ``controls'', correctly adjusting for patient characteristics should lead to the conclusion that target and control group have the same expected survival.  Note that, because the remaining control donor pool is still quite large,  \textit{very} close neighbors for most samples in the target group exist in the control pool. To gain more interesting insights on the effects of inter-and extrapolation biases, we therefore also investigate the performance of the different methods when we remove observations within $\delta_{min}$ average squared Euclidean distance across all features to any of the samples in the target group. }

\begin{wrapfigure}{r}{0.35\textwidth}
\vskip -0.1in
    \centering
\includegraphics[width=0.35\textwidth]{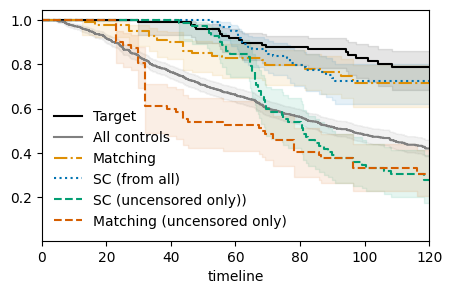}\vspace{-1cm}
    \caption{\textbf{Illustrating censoring bias.} \small Kaplan-Meier curves for $\delta_{min}=0.1$.} \vspace{-5mm}
 \label{fig:cens-illustration}
\end{wrapfigure}\textbf{\textbullet{ }Handling censoring.} Unlike the setup considered in the preceding theoretical sections, there is  censoring in the Rotterdam data. While this would not be problematic if the CAR assumption was fulfilled, it seems that censoring is actually \textit{not} at random here, because longer times are naturally more likely to be censored as follow-up is limited (indeed, only 50-70\% of individuals experience an event by 120 months). Constructing matched survival curves from only uncensored individuals, be it through matching or through synthetic control, therefore leads to downward bias as we can see in \cref{fig:cens-illustration}. Instead, we therefore apply the synthetic control weights to both the censored times $T^-_j = min(T, C)$ \textit{and} the event indicators $E_j=\mathbf{1}\{T < C\}$ and heuristically censor synthetic control units with $\sum^m_{j=1}w_jE_j < 0.5$. In \cref{fig:cens-illustration} we observe that this clearly improves upon using uncensored individuals only, so we use this approach throughout the section.

\begin{wrapfigure}{r}{0.6\textwidth}
\vskip -0.1in
    \centering
\includegraphics[width=0.6\textwidth]{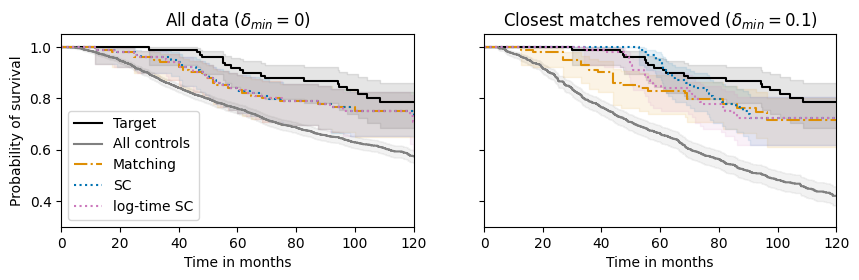}\vspace{-1cm}
    \caption{\textbf{Illustrating differences in shapes of (Kaplan-Meier) control survival curves.}}\vspace{-5mm}
 \label{fig:kmcurves}
\end{wrapfigure}\textbf{\textbullet{ }The shapes of survival curves.} In  \cref{fig:kmcurves} we illustrate the effect of the different approaches when used to create marginal survival curves. We make multiple interesting observations: First, when there are very close matches in the control data ($\delta_{min}=0$, left panel), all methods essentially perform the same -- this because there are nearest neighbor matches that are simultaneously an (almost) perfect synthetic control unit. When not ($\delta_{min}=0.1$, right panel), we observe that, as expected, standard SC tends to overestimate time more often while log-time SC tends to underestimate. Also, both versions of SC result in survival curve shapes that are qualitatively less close to the target distribution in shape than matching -- as expected, both SC distributions underestimate the occurence of early events.

\begin{wrapfigure}{r}{0.3\textwidth}
\vskip -0.15in
    \centering
\includegraphics[width=0.3\textwidth]{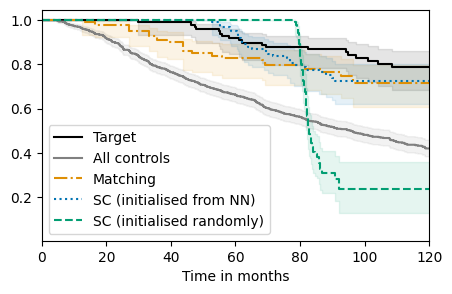}\vspace{-1cm}
    \caption{\textbf{SC Initialization effects.} \small Kaplan-Meier curves for $\delta_{min}=0.1$.} \vspace{-5mm}
 \label{fig:initbias}
\end{wrapfigure}\textbf{\textbullet{ }The effect of initialisation.} In \cref{fig:initbias}, we illustrate that this effect is exacerbated when we change the initialisation scheme to initialise SC with random weights: in this case, because solutions are not unique, SC tends to converge to a solution with many more contributors, so that survival times are indeed much too concentrated around their mean as expected. Throughout, we therefore initialise SCs from the nearest neighbor (NN) match.

\begin{figure}[h]
\subfigure[KS statistic]{
    \includegraphics[width=.45\textwidth]{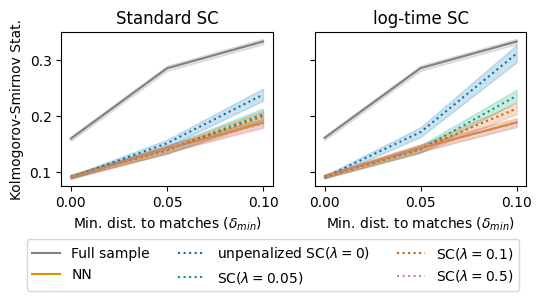}}
    \label{fig:mae}
\subfigure[MAE of predicting RMST]{
   \includegraphics[width=.45\textwidth]{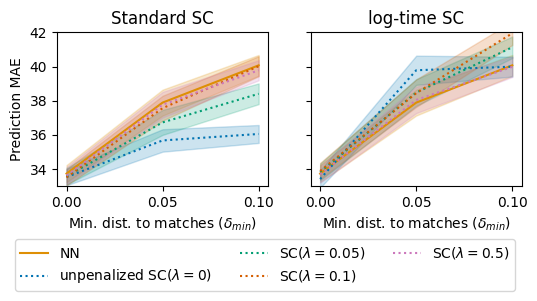}}
    \label{fig:rmst}\vspace{-.2cm}
\caption{\textbf{Performance of different methods by $\delta_{min}$. }\small Averaged over 20 experiments, 2SEs shaded.}
    \label{fig:res-sim}\vspace{-.5cm}
\end{figure}

\textbf{\textbullet{ }Trade-off between performance in terms of time prediction and survival curve estimation.} In \cref{fig:res-sim}, we now compare SC and matching more quantitatively in terms of (a) a metric for how close the survival curve is to the target survival curve (we use the Komolgoriv-Smirnov (KS) statistic for this) and (b) a metric capturing how close the individual predictions of survival time are (we use the mean absolute error (MAE) of predicting the Restricted Mean Survival Time (RMST) for this). We find that, as expected, NN matching does better at mimicking the survival curve shape -- but SC can do better at predicting the individual times. We also find that log-time SC does substantially worse than standard SC here, which might be due to any of the following reasons: i) underestimation of times exacerbates the exisiting observational bias here, ii) the signal-to-noise ratio might be low and iii) the true survival time might not actually follow an AFT model. 

\textbf{\textbullet{ }Effects of regularization. } Finally, we investigate whether incorporating weight penalty $\lambda_{var}$ can help with the observation above. In \cref{fig:res-sim}, we find that using small regularization penalties indeed leads to control cohorts with properties between the extremes of NN matching and SC: they perform better in terms of distribution metrics than SC, and better in terms of prediction metrics than matching -- but not as good as the best at either.  This indicates that the best strategy to use in practice depends on the objective: If getting the individual time predictions right is of most importance and there are no close matches in the data, then one may want to rely on synthetic control methods directly. If only the marginal distribution is of interest, then using matching alone may suffice. If, however, a practitioner is interested in trading off the two objectives, then the use of the regularization penalty can allow to find a place on the Pareto frontier between the two extremes.

\subsection{Experiment 2: Exploratory analyses of different treatment groups in real data}
\begin{figure}[t]
\subfigure[Chemotherapy]{
\includegraphics[width=.48\textwidth]{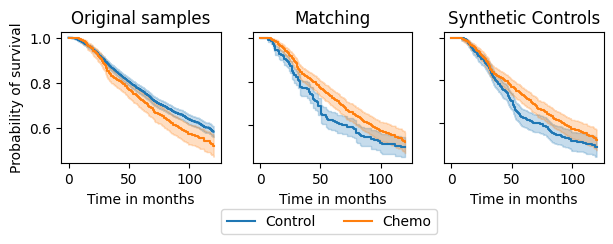}}
    \label{fig:chemo}
\subfigure[Hormonal Treatment]{
   \includegraphics[width=.48\textwidth]{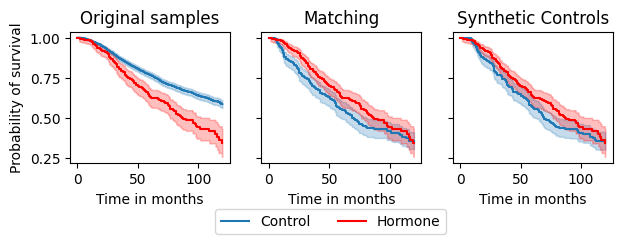}}
    \label{fig:hormon}\vspace{-.2cm}
\caption{\textbf{Exploratory results adjusting for biases in real data.} Kaplan-Meier survival curves for the original treated patients and control cohorts created using different methods.}\vspace{-.5cm}
    \label{fig:real-surv-curves}
\end{figure}

Plotting marginal survival in the real data as in \cref{fig:underlyingsurv} appears to -- counter-intuitively -- show that, on average, individuals who received a hormonal treatment or chemotherapy have shorter survival than those patients that received neither. We now investigate whether this unexpected observation can be \textit{explained away} by adjusting for observed differences between the two cohorts through creating real and synthetic matches in \cref{fig:real-surv-curves}. For both treatments, we observe that creating matched cohorts through either NN-matching or standard SCs (we drop log-time SC here due to consistently worse performance in \cref{sec:exp1}) leads to qualitatively the same conclusion and indeed results in reversal of the originally  counterintuitive finding: once we create a control group more aligned with the treated group, treated individuals indeed \textit{no longer} seem to live shorter -- to the contrary, there now appears to be some evidence that treatment increases the expected survival time. This shift in survival curves once patient characteristics are adjusted for is entirely expected -- only node positive patients actually received either therapy in the data \citep{foekens2000urokinase} -- and the evidence for treatment benefit, once groups are more matched in terms of patient characteristics, is in line with the original findings of the study where therapy indeed had a positive effect on survival \citep{royston2013external}.

\section{Conclusion} 
\squish{In this paper, we investigated the possibility of using synthetic control methods in survival analyses, and discovered multiple biases arising uniquely in this context. We showed that these are rooted in differences in both (i) commonly assumed DGPs and target parameters of interest in survival analyses relative to standard settings in economics, and (ii) differences in distributions of real and synthetic control units. We also highlighted that regularization schemes interpolating between matching and synthetic control methods may constitute a promising avenue to control some of these biases.}

We hope that future methodological work will build upon this initial evidence to develop more sophisticated approaches to overcome the challenges investigated in this paper. Further, we put our focus on a specific family of survival models (AFT models) in the theoretical part of this paper. It would therefore be another interesting avenue for future research to investigate whether our analyses can be extended to the proportional hazards family of survival models.  Finally, we chose to focus on comparing synthetic controls to matching, but did not investigate the possibility of imputing missing counterfactual outcomes in \textit{other} ways, e.g. using model-based approaches. Additionally comparing to model-based imputation of control survival times, using e.g. AFT- or Cox-regression models fitted using the control data, will add the potential for model misspecification as another dimension to the comparison, and may thus lead to further interesting tradeoffs (in addition to trading off interpolation and extrapolation biases as in the current comparison). Investigating the (dis)advantages of synthetic control methods relative to other possible approaches in the survival context could thus also be a fruitful direction to explore in future work.

\newpage
\bibliography{references}
\newpage
\appendix
\section{On the necessity of linearity assumptions for synthetic control methods}\label{app:linearity}
In this section, we briefly discuss whether linearity assumptions on the outcome-generating process, as in \cref{eq:sc_unb}, can sometimes be circumvented while retaining the ability of synthetic control methods to succeed. Indeed, 
\cite{shi2022assumptions} show that linearity assumptions on the outcome-generating process are not strictly necessary for synthetic control methods \textit{if one makes a different assumption} on the DGP (which, as we discuss at the end of this section, is not applicable in our survival setting). 

\cite{shi2022assumptions} make use of  a \textit{more fine-grained model assumption} than is usually the case in the synthetic control literature. Where classical synthetic control settings usually just assume the goal to estimate $\mu_{j, t}$ for units $j$ and times $t$, \cite{shi2022assumptions} make the additional assumption that every macro-unit $j$ is an \textit{average} of individual units $l$, i.e. that $\mu_{j, t}=E_i[\mu_{l, j, t}]$. In this case,  they show that ``linearity is a consequence of the fact that expectation is a linear operator'' \citep[p.2]{ shi2022assumptions} -- i.e. linearity is an automatic feature of the DGP of the \textit{aggregate units} even if the DGP of the individual units is not linear. They conclude that this is why synthetic control applied to aggregate units can work even if the DGP of the individual units are nonlinear. Conversely, they show that even within this setting, if the aggregation across individuals is \textit{not linear} (e.g. if using the group median instead of the group mean to construct aggregate units from the individual units), synthetic controls will be biased. Further, they also discuss that, if one wanted to include covariates beyond lagged outcomes, the ``use of auxiliary covariates requires us to assume that the underlying data generating process is linear'' \citep[p. 7]{shi2022assumptions}. In summary, \cite{shi2022assumptions} thus show that one may sometimes be able to remove the (parametric) linearity assumption usually used in synthetic control papers by instead assuming a fine-grained data-generating model that \textit{linearly aggregates unobserved individual units into macro-units} by averaging them. That is, some form of assumed linearity appears to remain crucial for successful application of synthetic control methods.

\paragraph{Why the fine-grained model is less applicable in survival analyses.} The additional assumption on the decomposition of aggregate units into individuals made  in the fine-grained model of \cite{shi2022assumptions} can be very reasonable within the economic applications considered in most of the original synthetic control literature: in such applications, one is often analyzing macro-level data from e.g. states or countries, which are indeed usually reported in aggregated form across counties, cities or even individuals. In our medical setting, however, the fine-grained model is not appropriate, as the final outcomes we consider are patient-level survival outcomes which are by construction already at an individual level; no further disaggregation is possible. Thus, the arguments of \cite{shi2022assumptions}, which rely on linear aggregation across individual units in the generation of observed outcomes, do not apply to our setting and can hence not help in mitigating the biases arising due to non-linearity present in survival DGPs.

\section{Experimental details}\label{app:exp}

\paragraph{Sampling scheme.} For the experiments in \cref{sec:exp1}, the target and control groups are constructed from the real data through the following biased sampling scheme: We first fit a cox proportional hazards model using all covariates on the full $n=2091$ patients, and use it to predict the expected median survival time $T^{med}_i(X_i)$ of each patient, which we normalize across the sample to give one value $Z_i = normalize(T^{med}_i(X_i))$ per patient. Second, we then split the $n$ patients randomly into two samples: with probability $.1$ they become part of the target pool $\mathcal{I}_t$, and with probability $.9$ they become part of the control pool $\mathcal{I}_c$. Third, from the target pool $\mathcal{I}_t$, we then create a biased target sample with higher expected survival times by sampling patients into the sample $\mathcal{S}_t$ with probability $p_i=expit(3*Z_i)$. Finally, when $\delta_{min}>0$, we then remove all close neighbors of patients in $\mathcal{S}_t$ from $\mathcal{I}_c$ to create the control sample $\mathcal{S}_c$, where the squared distance for each patient $j$ in the control sample for each patient $k_*$ in the target sample is calculated from the normalized d-dimensional covariate vectors as $\delta_{jk_*}= \frac{1}{d}\sum^D_{d=1}(X^d_j - X^d_{k_*})^2$. 

\paragraph{Metrics.} The Komolgorov-Smirnov statistic measures the distance between two CDFs $F(x)$ and $G(x)$ as $sup_x |F(x) - G(x)|$. We approximate this from the target samples and the different control groups by first computing the empirical survival functions $\hat{S}^t$ and $\hat{S}^c$ using the Kaplan-Meier estimator, and then computing $max_{T \in \mathcal{S}_t} |\hat{S}^t(T) - \hat{S}^c(T)|$. 

The restricted mean survival time (RMST) relative to time $t_{end}=120$ months is defined as $T^{RMST}_i=min(T_i, t_{end})$. To evaluate individual predictions, we compute the mean absolute error between predicted and observed RMST for each unit in the target sample, where we can only use individuals $i$ for which both observed and predicted time are not censored before $t_{end}$ : \begin{equation}
    MAE = \frac{\sum_{i \in \mathcal{S}_t} \mathbf{1}\{i: (E_j=1 | T^-_j>t_{end}) \& (\hat{E}_j=1 | \hat{T^-}_j>t_{end})\} |\hat{T^{RMST}}_j - {T^{RMST}}_j|}{\sum_{i \in \mathcal{S}_t} \mathbf{1}\{i: (E_j=1 | T^-_j>t_{end}) \& (\hat{E}_j=1 | \hat{T^-}_j>t_{end})  \}}
\end{equation}

\end{document}